\documentclass[conference]{IEEEtran}
\IEEEoverridecommandlockouts{}
\usepackage{cite}
\usepackage{amsmath,amssymb,amsfonts}
\usepackage{algorithmic}
\usepackage{graphicx}

\usepackage{textcomp}
\usepackage{xcolor}
\usepackage{booktabs}
\usepackage{multirow}
\usepackage{multicol}
\usepackage{url}
\def\BibTeX{{\rm B\kern-.05em{\sc i\kern-.025em b}\kern-.08em
    T\kern-.1667em\lower.7ex\hbox{E}\kern-.125emX}}
\begin{document}
\title{TRec: Sequential Recommender Based On Latent Item Trend Information \\
\thanks{Can Wang is the corresponding author}
\thanks{This work is supported in part by Griffith Industry Collaborative Grant}
\thanks{This work is supported in part by the National Key Research and Development Program of China under Grant 2017YFE0117500}
\thanks{This work is supported in part by the Natural Science Foundation of the Higher Education Institutions of Jiangsu Province(Grant No.17KJB520028), Tongda College of Nanjing University of Posts and Telecommunications(Grant No.XK203XZ18002)}
}

\author{\IEEEauthorblockN{1\textsuperscript{st} Ye Tao}
	\IEEEauthorblockA{\textit{School of ICT} \\
		\textit{Griffith University}\\
		Gold Coast, Australia \\
		ye.tao2@griffithuni.edu.au\\}
	\and
	\IEEEauthorblockN{2\textsuperscript{nd} Can Wang}
	\IEEEauthorblockA{\textit{School of ICT} \\
		\textit{Griffith University}\\
		Gold Coast, Australia \\
		can.wang@griffith.edu.au}
	\and
	\IEEEauthorblockN{3\textsuperscript{rd} Lina Yao}
	\IEEEauthorblockA{\textit{School of Computer Science} \\
		\textit{University of New South Wales}\\
		Sydney, Australia \\
		lina.yao@unsw.edu.au}
	\and
	\IEEEauthorblockN{4\textsuperscript{th} Weimin Li}
	\IEEEauthorblockA{\textit{School of Computer Engineering and Science} \\
		\textit{Shanghai University}\\
		Shanghai, China \\
		wmli@shu.edu.cn}
	\and
	\IEEEauthorblockN{5\textsuperscript{th} Yonghong Yu}
	\IEEEauthorblockA{\textit{Tongda College} \\
		\textit{Nanjing University of Posts and Telecommunications}\\
		Nanjing, China \\
		yuyh@njupt.edu.cn}
}

\maketitle

\begin{abstract}
	Recommendation system plays an important role in online web applications.
	Sequential recommender further models user short-term preference through exploiting information from latest user-item interaction history.
	Most of the sequential recommendation methods neglect the importance of ever-changing item popularity. We propose the model from the intuition that items with most
	user interactions may be popular in the past but could go out of fashion in recent days.
	To this end, this paper proposes a novel sequential recommendation approach dubbed TRec, TRec learns item trend information from implicit user interaction history and incorporates item trend information into next item recommendation tasks. Then a self-attention mechanism is used to learn better node representation.
	Our model is trained via pair-wise rank-based optimization.
	We conduct extensive experiments with seven baseline methods on four benchmark datasets,
	The empirical result shows our approach outperforms other state-of-the-art methods while maintains a superiorly low runtime cost.
	Our study demonstrates the importance of item trend information in recommendation system designs, and our method also possesses great efficiency which enables it to be practical in real-world scenarios.

\end{abstract}


\section{INTRODUCTION}
As the web-based entertainment and e-commerce application becoming popular, 
there has grown a urgent need for both users and web service providers to filter the overloaded information. 
In most of the time users are inundated with these options. 
Recommender system is born to address such information overload problems. \cite{FANG2019}.

Collaborative filtering-based (CF) \cite{Sarwar2001} approaches and content-based (CB) approaches \cite{Pazzani2007} are two groups of mainstream traditional recommendation methods.
More recently, the latent factor based models such as MF(matrix factorization) \cite{Koren2015} and FPMC(Factorizing Personalized Markov Chain)\cite{Rendle2010} have replaced CF and CB for its better prediction accuracy and the lower runtime cost.
Factorization-based methods are able to conveniently model the representations of user long-term preference with the user index.
However, many works \cite{Jannach2015,Zhang2019} have pointed out that the matrix factorization based approaches fail to obtain an appropriate representation for the user's ever-changing taste.
To solve this issue, recent works incorporate the user short-term preference into the traditional user latent vector \cite{Zhang2018}.
User short-term preference can be derived from the recent user-item interaction history.
Leveraging upon the mixed user long-term and short-term preferences,
the prediction model is therefore able to comparably give a higher weight score on the recent user interaction behaviors.
Nevertheless, most of these works neglect the similar phenomenon:
the extent to which the items are accepted and loved by the mainstream users is in a state of flux.
For example, we take kung-fu movie as an instance, kung-fu or martial art movies are popular in 1980s \cite{Fu2001}. As the trend of enthusiasm for those kung-fu movies ebbs, however, even the former fans of such a genre might be less likely to watch a latest released kung-fu movie.
It actually has nothing to do with the quality of the item itself. That is to say, we need to take this ever-changing popularity of item into account.
Recent recommender systems tend to neglect such a shifting item popularity. For example,
based on the former kung-fu lover's interaction history, a traditional factorization-based recommender might give a high prediction score on a target kung-fu movie. As a result, however, we can see that both the classical matrix factorization approaches \cite{Jannach2015,FANG2019,Zhang2019} and the recent deep learning based approaches fail to find a way to
model items' changing popularity. However in the context of sequential recommender systems, the item is still regarded as a static representation. It thus neglects latent trend of an item.
Aiming to deal with the aforementioned problems, our work contributes in the following aspects:
\begin{itemize}
	\item We propose a novel sequential recommender framework: TRec (Trendy Recommender). Our approach takes the target item's latest activity history into account, as well as the user's temporary and long-term preferences.
	\item We utilize the self-attention mechanism for modeling the representation of item trend information. Our model demonstrates the excellent learning ability on the implicit interaction scenarios.


	\item Comprehensive experiments are conducted in comparison with several state-of-the-art recommender frameworks.
	      Our approach shows promising results over the baseline models. The improvement rate ranges from 6\% to 11\% in different metrics, when compared with the state-of-the-art models in terms of accuracy on most popular datasets.
\end{itemize}


\section{RELATED WORK}
In this section we will introduce related work that falls into sequential recommendation category, in the perspective of traditional and deep learning approaches.
\subsection{Sequential Recommendation}
Sequential Recommendation System (SRS) is a hot research topic in the research area of recommendation systems \cite{FANG2019}.
The main-stream methods bifurcate into two categories: traditional non-neural approaches, and neural-based or deep-learning based approaches.
\newline
\textbf{Traditional Approaches}\hspace{0.2cm}Most traditional recommender systems utilize collaborative filtering based methods. Specifically, they incline to utilize a user's historical interactions to learn her static preference with the assumption that all user-item interactions in the historical sequence are equally important\cite{FANG2019}.
In real world scenario, user behavior is usually not determined merely upon her all-time preference, a basic observation is that the interest of user changes all the time.
Under this premise the sequential recommendation is brought into beings.
SRS uses user's historical interaction to construct sequences. Then these sequences are modeled and used to predict the possibility of next item which user would like.
Earlier sequential recommender approaches employ Markov Chain\cite{Davidson2010} and session-based KNN\cite{Jannach,He2017},which fall short in modeling user long-term preferences.
Factorization-based methods such as matrix factorization\cite{Koren2015} and its variant are widely used in industry for its fast speed and acceptable performance.

\textbf{Deep Learning Approaches}\hspace{0.2cm} In recent days deep neural networks techniques are widely absorbed into sequential recommendation.
RNNs, CNNs and attention mechanism are the mainly adopted approaches for DL based sequential recommendation.
RNNs architecture have been well exploited in sequential recommendation domain, Hidasi et al.\cite{Hidasi2016} proposed GRU4Rec which is the
first to have introduced RNNs into sequential recommendation, however GRU4Rec fails to take user information into modelling.
Due to potential limitation on sequence length and expensive computing costs, RNNs based models were surpassed by CNNs based framework
and attention mechanism based framework.
Caser\cite{Tang2018} views the embedding matrix of L previous items as an image therefore convolution operation could be done for prediction.
Yuan et al. \cite{Yuan2019} proposed NextItNet which utilized residual block CNNs architecture on sequential recommendation tasks.
Liu et.al\cite{Liu2018a} proposed a method where using vanilla attention for calculating the scores of items in a given sequence.
Zhang et.al\cite{Zhang2018} proposed AttRec which integrated self-attention mechanism into sequential recommendation.
As most sequential based methods make prediction via dot product between user and item vector representation but few make effort on improving item representation learning, therefore
seeking improvement in item representation could be a promising research direction.

\section{Model Overview}
\addtolength{\dbltextfloatsep}{-10pt}
\addtolength{\textfloatsep}{-15pt}
\addtolength{\intextsep}{-8pt}
\addtolength{\floatsep}{-15pt}
\begin{figure*}[htbp]
	\centerline{\includegraphics[width=\linewidth]{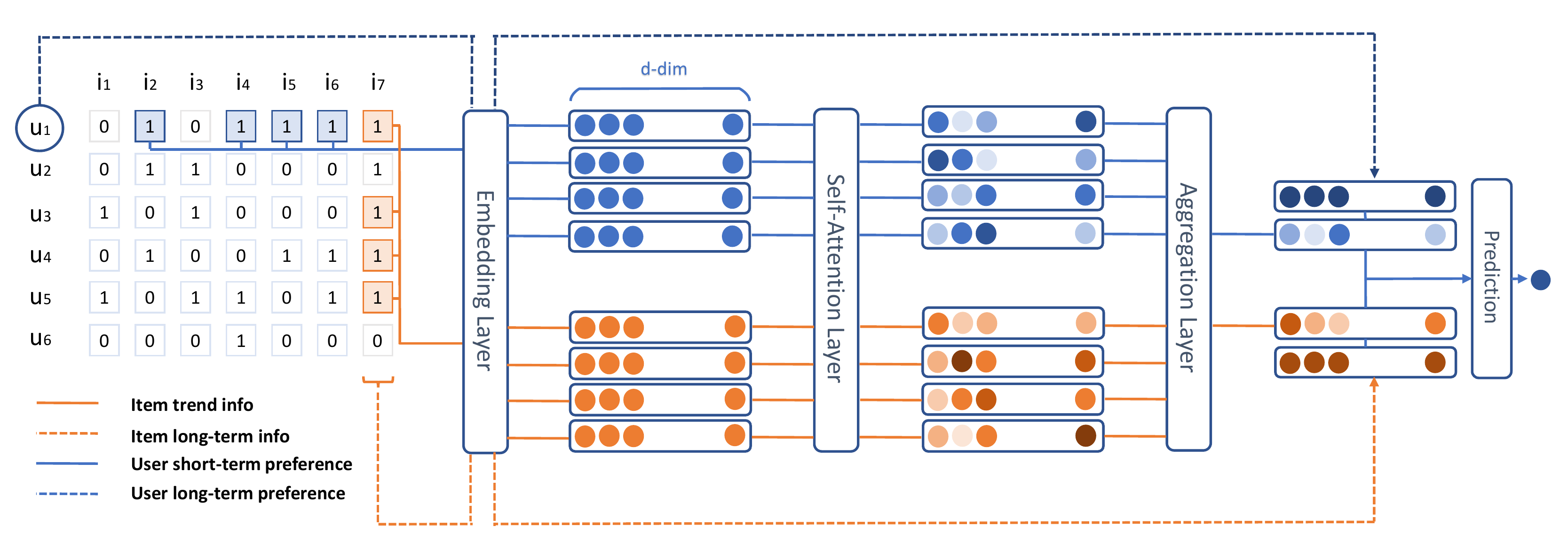}}
	\caption{TRec:Overview of the architecture. The blue and orange squares of the user-item interaction matrix indicates user interaction sequence
	and top-k recent users respectively. In this case top 4 recent users is selected for item trend information modeling. The blue and orange dashed
	line modeled from their indices denote user long-term preference and item long-term trend respectively. The user long-term representation and
	 item long-term representation are gathered together with user short-term representation and item trend information after aggregation step for further prediction jointly.}
	\label{overview}
\end{figure*}

In this section, we present a brief outline of our proposed model to explain our method in a nutshell. 
The inputs of our model comprises four part: the long-term user preference and item information, the short-term user preference, and the item trend information.
As Fig. \ref{overview} shows, before the recommender system makes a recommendation item list to a user, the recent interaction trajectory of that user is used as an input to our model for quantifying short-term user preference,
which is colored in blue solid line. Here the interaction of a user can be seen as an explicit or implicit behavior committed to an item.
Then The item trend information, which is colored in orange dashed line, is learned through the top-k recent user list of whom have interacted with one particular item, and this is used as an input for the item trend representation learning.
The long-term user preference and long-term item information are colored in blue dashed line and orange dashed line respectively. The representation of them are based of their index and learned through the training process.
These four inputs are fed into the embedding layer which output the d-dimension embeddings of them.
An embedding is a d-dimensional vector which is constructed as a representation for a user or a item. 
Then we take advantage of a self-attention layer in order to catch more precisely representation of embedded inputs.
Inspired by the recent NLP(Natural Language Processing) progress, self-attention\cite{Vaswani2017} mechanism is employed in our model to provide a clear representation that catches the inner correlation of features within a user embedding or an item embedding. 
A semantic explanation of the modelling process of item trend information can be viewed in a way that the interrelationships of users contribute to the item popularity, which essentially distinguishes our method from other sequential recommender models.
At last, the re-represented embedding matrix of inputs are going to be aggregated through the aggregation layer in order to fit the shape for the following dot-product operation, 
and in the prediction layer our model gives a prediction score for each item that user have not interacted. According to the descending ordered list, items
with the top-k highest prediction scores are selected as the recommendations for this particular user.



\section{TRec: Methodology and Model}
In this section, we go through each component of our model, as shown in Fig. \ref{overview}. Firstly, we give a formulated definition on the sequential recommendation, and then we introduce the concept of implicit interactions. In the following, we explain the definition of item trend information and user short-term preference, and elaborate how they are
modeled through embedding layer and self-attention layer, and the relationship between those two layers. Next, we demonstrate how these information together are used jointly to yield a prediction score.
At last we present the time complexity analysis result.


\subsection{Formulation of Our Model}
At first, to be consistent, we would like to clarify the notation convention used in this paper as follows: scalars are denoted in plain typeface, a lowercase bold letter denotes a vector, and a uppercase boldface denotes a matrix.
\par
A formulation description of our model is given as follows. Assuming we have the user-item interaction list of users \(u\) which is denoted as \(L^u=\{v_k|k\in H_u\} \),
Similarly, the item list \(L^v=\{u_k|k\in H_v\} \) is the set of users who have interacted with the particular item \(v\), where \(H_v\) is the set of item indices that denotes interactions.
If our model is regarded as a function \(\mathcal{F}(\cdot )\) with all the parameters denoted as \(\Theta\),
then the sequential recommendation task can be formulated as follows:
\begin{equation}
	R = \mathcal{F} [(L^u,\mathbf{L^v}); \Theta],
\end{equation}
where \(R\) is the list of items to be recommended to \(u\), \(\mathbf{L^v}\) is the matrix of \(L^v\) for all the items in the whole dataset.
\par
Table \ref{notation} displays the most frequently used notations in this paper.
\begin{table}[htbp]
	\begin{center}
		\caption{List of notations}
		\label{notation}
		\begin{tabular}{ll}
			\toprule
			notation                    & meaning                                                        \\
			\midrule
			$L^u$                       & interaction history list of $u_i$                              \\
			$L^v$                       & set of users have interacted with $v_i$                        \\
			$H_u$                       & interacted item indices of $u_i$                               \\
			$H_v$                       & indices of users interacted with $v_i$                         \\
			$L_p^u$                     & sequence of items $u_i$ recently interacted                    \\
			$L_q^v$                     & sequence of top-q users interacted with $v_i$                        \\
			$\mathbf{u^+}$              & short-term preference of $u_i$                                 \\
			$\mathbf{v^+}$              & trend information of $v_i$                                     \\
			$\mathbf{L_q^v},\mathbf{L_p^u}$ & embeddings of $L_q^v$ and $L_p^u$                                  \\
			$\hat{r}^i_j$               & output of score of $v_j$ w.r.t $u_i$                           \\
			$\Theta$                    & parameters to be learned                                       \\
			$d$                         & dimension of embedding                                         \\
			$p,q$                       & the lengths of $L_p^u$ and $L_q^v$                                 \\
			$\omega,\alpha,\beta$       & $\omega$ indicates the proportion to which the temporary       \\
			                            & and long-term user preference contributes,$\alpha$ and $\beta$ \\
			                            & indicates to what extent the item trend information            \\
			                            & should be taken into account.                                  \\
			\bottomrule
		\end{tabular}
	\end{center}
\end{table}

\subsection{Implicit Interaction}
Our proposed approach addresses the next item prediction problem with the implicit interaction histories. An implicit interaction
is a style of user interaction that does not require the explicit behaviors but still shows the user preference. 
For example, a click, or a purchase falls into the category of implicit interaction, 
while the direct rating behavior of a movie belongs to the explicit interaction. 
We model the implicit interaction using a two-value system, where 1 indicates the user has interacted with a item, 
and 0 indicates the user dislikes or has not interacted with a item yet. For example, as Table \ref{iti example} shows, \((u_1,v_1)\)
equals 1 means user \(u_1\) had interacted with item \(v_1\).

\subsection{Inputs and Embedding Layer}
\noindent\textbf{User Short-term Preference Modeling}\label{U+}\hspace{0.2cm}
Let us assume for a interaction history sequence \(L_u\) of user $u$, an recent user-item sequence \(L_p^u\) with the length \(p\) is extracted for short-term preference modelling, which is represented as
\begin{equation}
	L_p^u=\{ v_1,v_2,\ldots,v_{p} \}.
\end{equation}

Here we take Table \ref{iti example} as an example, in this case we take \(p=4\) for convenience.
We can see that the user \(u_1\) has recently watched \(v_1,v_4,v_5,v_7\) those four movies. Then, we have \(L_4^{u_1} = \{v_1,v_4,v_5,v_7\}\).

The embedding vector of each item that the user \(u\) has interacted with, is obtained by looking up through an item embedding matrix which is initialized with random values and will be updated in the learning process. Thus, the embedding matrix of \(L_p^u\) can be denoted by
\begin{equation}
	\mathbf{L_p^u} = (\mathbf{v_1},\mathbf{v_2},\ldots \mathbf{v_p}),
\end{equation}
where \(\mathbf{L_p^u}\in\mathbb{R}^{d\times{p}}\), and \(d\) denotes the dimension of embedding of items.
\newline
\textbf{Item Trend Information Modeling}\hspace{0.2cm}
\begin{table}[htbp]
	\begin{center}
		\caption{An Instance of Item Trend Information}
		\label{iti example}
		\begin{tabular}{ccccccccll}
			\toprule
					 & \(v_1\)  & \(v_2\)   & \(v_3\) & \(v_4\)&\(v_5\) & \(v_6\) & \(v_7\) & \(v_8\)\\
			\midrule
			\(u_1\)   & 1 & 0 & 1 & 1& 1 & 0 & 1 & \textbf{1}\ldots(Apr-3-12:30)     \\
			\(u_2\)   & 0    & 0   & 1 & 1 &0&1&0&0  \\
			\(u_3\) & 1 & 1 & 0 & 1 &1&0&0&\textbf{1}\ldots(Apr-1-22:23)      \\
	 		\(u_4\)   & 0 & 0 & 1 & 0&1&0&0&\textbf{1}\ldots(Feb-2-12:30)      \\
			\(u_5\)   & 1 & 1 & 0 & 1 &1&0&0&0      \\
	 		\(u_6\)   & 0 & 0 & 0 & 0&1&0&0&\textbf{1}\ldots(Feb-1-2:20)      \\
			\(u_7\) & 1 & 0 & 0 & 0 &1&0&0&1\ldots(Feb-1-8:32)      \\
			\(u_8\) & 1 & 1 & 0 & 1 &1&0&0&1\ldots(Jan-12-2:30)      \\
	 		\(u_9\)   & 0 & 0 & 1 & 0&1&0&0&1\ldots(Jan-11-11:08)      \\
	 		\(u_{10}\)   & 0 & 0 & 1 & 0&1&0&0&1\ldots(Jan-3-4:20)      \\
			\bottomrule
		\end{tabular}
	\end{center}
\end{table}
Assuming for the item \(v\), the list of users who have interacted with \(v\) is denoted as $L^v=\{u_k|k\in H_v\}$,
where \(H_v\) is the set of indices that denotes the index of users who have ever interacted with the item \(v\). We extract the latest top \(q\) elements of the set \(L^v\) for modelling the representation of the trend information of this item. Accordingly, we denote it as
\begin{equation}
	L_q^v = (u_1,u_2,\ldots,u_q).
\end{equation}
Let us take Table \ref{iti example} as an example. As it shows, the elements in the last column valued 1 indicate that those users have interacted with \(v_8\). Here in the last column, each interaction record is paired with a time-stamp. If we take \(q=4\) in this case, it  means that the top 4 recent interactions are used for the item trend information modeling. For convenience, the last column is sorted in the descending order in terms of time-stamp. If we want to select the top 4 records as the latest interaction history, as we boldface in Table \ref{iti example}, we then get the corresponding user list \(L_4^{v_8} = (u_1,u_3,u_4,u_6)\).
Similar to what we do in user short-term preference modelling , we look up through a embedding layer to obtain the embedding matrix of \(L_q^v\):
\begin{equation}
	\mathbf{L_q^v} = (\mathbf{u}_1,\mathbf{u}_2,\ldots,\mathbf{u}_{q}),
\end{equation}
where \(\mathbf{L_q^v} \in \mathbb{R}^{d\times q}\).
\newline
\textbf{Long-term Modeling for Users and Items}\hspace{0.2cm}
We denote the long-term representation of the user $u$ as \(\mathbf{u}_i\). Accordingly, the long-term representation of the item $v$ is denoted as \(\mathbf{v}_i\). In our approach, the long-term representations for users and items are two low-rank vectors with \(d\) dimensions, which are obtained through a look-up operation from separated embedding matrices.
Here, \(d\) is a hyper-parameter which is specified by the user.
For example, the index \(i\) of one particular user \(u_i\) is used to get a corresponding embedding vector \(\mathbf{u}_i\).
The parameters of the embedding are updated via the back-propagating process in each training epoch.
\subsection{Self-Attention and Aggregation Layer}
\noindent\textbf{Self-Attention Layer}\hspace{0.2cm}The embeded user-item interaction list \(\mathbf{L_p^u}\) and item trend information \(\mathbf{L_q^v}\) are fed into a self-attention layer
for better representation modelling. Self-attention function \cite{Vaswani2017} is defined as follows:
\begin{equation}
	attention(\mathbf{Q,K,V})=softmax(\frac{\mathbf{QK}^T}{\sqrt{d}})\mathbf{V},
\end{equation}
where \(\mathbf{V}\) is the input matrix, and \(\mathbf{Q,K}\) stand for \(query\) and \(key\) respectively, which are the matrices
mapped from input \(\mathbf{V}\) with the weight matrices \(\mathbf{W_Q}\)
and \(\mathbf{W_K}\), respectively, where \(\mathbf{Q}=ReLu(\mathbf{W_QV})\), \(\mathbf{K}=ReLu(\mathbf{W_KV})\).
In addition, the product is fed into a \(ReLu\) non-linear activation
function.
ReLu (Rectified Linear Unit) \cite{Nair2010} is a widely used non-linear activation function in deep learning.
It can be formalized as follows:
\begin{equation}
	ReLu(x) = max(0,x).
\end{equation}


On the other hand, \(\sqrt d\) is a scaling factor used to scale the dot product above, thus avoiding the gradient to end up becoming too small. 


At last, a softmax layer is used to map the output to a range between 0 and 1, which can be viewed as the affinity of each feature in the matrix. The shape of the output matrix remains the same as input.
\newline
\textbf{Aggregation Layer}\hspace{0.2cm}
Aggregation layer is used to fit the shape of output of embedding matrix in order to be able to perform following dot-product operation.
In experiment section we analysed average aggregation and max aggregation, here we take average aggregation as an example.
We aggregate the matrix \(\mathbf{L_p^u}\) and \(\mathbf{L_q^v}\) along each row as follows:
\begin{equation}
	\mathbf{u}^+ = (u^+_1,u^+_2,\ldots,u^+_d).
\end{equation}
The \(t\) th entry of $u^+_t$ is aggregated through:
\begin{equation}
	u^+_t = \frac{1}{p}\sum_{i}\mathbf{L}_{it}^u,
\end{equation}
where \(u^+\) indicates the short-term preferences.
Similarly, the latest top \(k\) item embedding matrix \(\mathbf{L_q^v}\) is fed into the self-attention layer as well.
We aggregate the output matrix \(\mathbf{L_q^v}\) along each column as follows:
\begin{equation}
	\mathbf{v}^+ = (v^+_1,v^+_2,\ldots,v^+_d).
\end{equation}
The \(t\) th entry of \(v^+_t\) is aggregated through:
\begin{equation}
	v^+_t = \frac{1}{q}\sum_{i}\mathbf{L}_{it}^v,
\end{equation}
where \(v^+\) indicates the item trend information.

\subsection{Prediction layer}
\noindent\textbf{Prediction Score Definition}\hspace{0.2cm}We define the prediction function as follows:
\begin{equation}
	\hat{r}^i_j = \omega \mathbf{u}_i(\mathbf{v}_j+\alpha\mathbf{v}^+_j)+(1-\omega)\mathbf{u}^+(\mathbf{v}_j+\beta\mathbf{v}^+_j),
\end{equation}
where \(\hat{r}^i_j\) is the next item to be recommended by our recommender algorithm, \(j\) is the index of the next item, \(i\) is the index of the current user to be recommended.
\(\omega\) indicates the proportion to which the short-term and long-term user preferences contribute,
\(\alpha\) and \(\beta\) specify to what extent the item trend information should be taken into account.
When \(\omega=1, \alpha=0\), it degrades to a matrix factorization based model \cite{Koren2015}.
\newline
\textbf{Loss Function Definition}\hspace{0.2cm}


Inspired by the BPR-Opt proposed by Rendal et.al \cite{Rendle2009}, which has been proved to provide a better ranking quality than the rating prediction based optimization methods in implicit interaction scenarios,
the loss function in our approach is defined as:
\begin{equation}
	\sum_{(u,i,j)\in D_s}\ln\sigma(\hat{r}^u_{i} - \hat{r}^u_{j},)-\lambda_\Theta||\Theta||^2,
\end{equation}
where \(D_s = \{(u,i,j|i\in L^u,j\in I \backslash L^u)\}\), and \(I\) denotes the items set.
The semantic explanation of \(D_s\) is that user \(u\) is assumed to prefer \(i\) over \(j\).
\(\hat{r}^u_{i,j}\) is the prediction score of the user \(u\) on the item \(i\) and \(j\),
\(\lambda_\Theta||\Theta||^2\) is the regularization term.
\newline
Let's take the movie recommendation as an example, our loss function aims to minimize the error so that a user has a greater possibility to watch the movie \(i\) over the movie \(j\).


\subsection{Time Complexity}
The time complexity of our model is mainly due to the self-attention layer. Assuming the single user interaction sequence to be \(L_q^v\), and \(L_p^u\) to be the sequence length of users who have interacted with one item recently.
The time complexity is \(\mathcal{O} ({|L_q^v|}^2d+|L_p^u|^2d)\). Thanks to the parallelizable nature of self-attention mechanism, our method has a relatively low sequential complexity of \(\mathcal{O}(1)\), compared with the RNN-based method such as GRU4Rec \cite{Hidasi2016} which has a sequential complexity of \(\mathcal{O}(n)\), since it has to wait for the output of time step \(t-1\).
The empirical study results evidence that our method runs the order of magnitude faster than other RNN-based and CNN-based methods.


\section{EXPERIMENTS}
In this section, extensive experiments are conducted to verify the effectiveness, sensitivity, and efficiency of our proposed model TRec using the item trend information. Accordingly, we would like to answer the following research questions:
\begin{itemize}
	\item RQ1: Is the item trend information useful in the sequential recommendation? If so, how do the key hyper-parameters affect the performance of our approach?
	\item RQ2: Does our approach outperform the state-of-the-art models?
\end{itemize}

\subsection{Experimental Settings}
\noindent\textbf{Dataset}\hspace{0.2cm}
We perform our experiments with four public datasets: three of them are the subsets of Amazon Custom Review Dataset: Luxury, Software and Digital. Amazon Custom Review Dataset \cite{Ni} is a widely accepted stable benchmark dataset for recommendation systems. The rest one dataset we adopt is MovieLens100K\footnote{https://grouplens.org/datasets/MovieLens/}, which is an online movie website for movie recommendation and building custom movie taste profiles.
The statistics of these four datasets are demonstrated in Table.\ref{dataset}, where the median in the header row means the median of rating counts per user. Based on this statistics, we conclude the facts as follows:

\begin{table}[htbp]
	\begin{center}
		\caption{Dataset Statistics}
		\label{dataset}
		\begin{tabular}{ccccc}
			\toprule
			Dataset  & Users  & Items   & Total Rating Counts & Median \\
			\midrule
			Luxury   & 12,369 & 416,425 & 536,554             & 32     \\
			ML100K   & 610    & 9,724   & 100,000             & 70.5   \\
			Software & 21,663 & 375,147 & 459,436             & 3      \\
			Digital   & 32,589 & 324,040 & 371,344             & 2      \\
			\bottomrule
		\end{tabular}
	\end{center}
\end{table}
\begin{itemize}
	\item MovieLens100K has the largest median of ratings per user (70.5 per user), which implies the users from movie websites behave more actively than those from the e-commerce shopping websites. Luxury subset comes a close second with 32 ratings per user in terms of median.
	\item Software and Digital subsets have a significantly fewer ratings per user. It suggests that users tend to buy fewer items of those categories.
\end{itemize}

\noindent\textbf{Evaluation Measures}\hspace{0.2cm}The dataset is split into three portions, 70 percent of original dataset is used for training,the rest is split into two parts: 20 percent for validation and 10 percent for testing.
We use two metrics to evaluate our approach versus other models in the way of Recall@K and NDCG@K. Recall@K indicates the ratio of users that have interacted with the items over
top-K recommended items, given by our prediction algorithm. NDCG (Normal Discounted Cumulative Gain) metric takes the position into account. It is often used to measure the effectiveness of web search engine algorithms or related applications.
NDCG measures the usefulness or gain of a document based on its position in the result list. The gain is accumulated from the top of the result list to the bottom, with the gain of each result discounted at lower ranks.


\noindent\textbf{Baseline Methods}\hspace{0.2cm} The following competitive methods which are popular in the sequential recommendation tasks are compared with our proposed approach.
\begin{itemize}
	\item \textbf{BPR-MF}\hspace{0.2cm} BPR-MF \cite{Rendle2009} is a well-known latent factor model for recommendation system, where rating are calculated through a dot product of two low-rank user and item latent vectors.
	\item \textbf{FPMC}\hspace{0.2cm} FPMC (Factorization Machines) \cite{Rendle2010} FPMC is a representative baseline for next-basket recommendation, which integrates the MF with first-order MCs. 
	\item \textbf{GRU4Rec}\hspace{0.2cm} GRU4Rec \cite{Hidasi2016} is the first model that applies RNN to the sequential recommendation, and does not consider a user's identity. The input of GRU4Rec is a set of items, and the embedded items matrix is fed into the stacked GRU layers for next item prediction. 
	\item \textbf{AttRec}\hspace{0.2cm} AttRec \cite{Zhang2018} utilizes the self-attention mechanism from NLP to infer the item-item relationship from the user historical interactions. It also takes user's transient interest into consideration. 
	\item \textbf{Caser}\hspace{0.2cm} Caser \cite{Tang2018} is a convolution based sequential recommendation model. It captures the high-order Markov chains via applying the convolution operations on the recent user interaction sequences.
	\item \textbf{HGN}\hspace{0.2cm}HGN (Hierarchical Gating Networks for Sequential Recommendation) \cite{Ma2019} adopts a hierarchical gating architecture to select what item features can be passed to the downstream layers from the feature and instance level. HGN outperforms several recent state-of-the-art models on different datasets.
\end{itemize}





\begin{table*}[htbp]
	\begin{center}
		\caption{Performance comparison over other baseline models. Best performance is boldfaced while scores underlined comes to a second place in performance. Improvement demonstrates the percentage of improvement of our model versus the best baseline model}
		\label{performance}
		\begin{tabular}{c|cccccccc|c}
			\toprule
			                           & Dataset       & BPR-MF     & FPMC     & GRU4Rec & AttRec             & Caser              & HGN                & TRec			 &Improvement            \\

			\midrule
			\multirow{4}{*}{Recall@10} & Amazon-Luxury & 0.1127 & 0.1219 & 0.1491  & 0.1746             & 0.1544             & \underline{0.1788} & \textbf{0.1972}	&10.29\%\\
			                           & MovieLens100K & 0.0712 & 0.0857 & 0.0889  & 0.0956             & 0.0982             & \underline{0.0965} & \textbf{0.1053} &9.12\%\\
			                           & Digital        & 0.0631 & 0.0660 & 0.0700  & 0.0704 			& 0.0723             & \underline{0.0752} & \textbf{0.0872} &15.96\%\\
			                           & Software      & 0.0812 & 0.0957 & 0.2009  & \underline{0.3216} & 0.2952             & 0.3205             & \textbf{0.3331} &3.58\% \\
			\midrule
			\multirow{4}{*}{Recall@20} & Amazon-Luxuxy & 0.2426 & 0.1691 & 0.1711  & 0.1918             & \underline{0.2094} & 0.1995             & \textbf{0.2310} &10.32\%\\
			                           & MovieLens100K & 0.1257 & 0.1492 & 0.1522  & 0.1529             & 0.1545             & \underline{0.1617} & \textbf{0.1735} &7.31\%\\
			                           & Digital       & 0.0712 & 0.0728 & 0.0811 & 0.0830				& 0.0851 	         & \underline{0.1074}& \textbf{0.1213} &12.94\%\\
			                           & Software      & 0.1320 & 0.1402 & 0.2518  & \underline{0.3816} & 0.3497             & 0.3791             & \textbf{0.4113} &7.89\%\\
			\midrule
			\multirow{4}{*}{NDCG@10}   & Amazon-Luxuxy & 0.0624 & 0.0675 & 0.0724  & 0.0842             & 0.0792             & \underline{0.0892} & \textbf{0.0981} &9.87\%\\
			                           & MovieLens100K & 0.0458 & 0.0495 & 0.0531  & 0.0618             & 0.0581             & \underline{0.0654} & \textbf{0.0699} &6.88\%\\
			                           & Digital		   & 0.0574 & 0.0625 & 0.0674  & 0.0790             & 0.0759             & \underline{0.0792} & \textbf{0.0855}&7.95\% \\
			                           & Software	   & 0.0524 & 0.0575 & 0.0624  & \underline{0.0700} & 0.0654             & 0.0689			 & \textbf{0.0746}&6.71\% \\
			\midrule
			\multirow{4}{*}{NDCG@20}   & Amazon-Luxuxy & 0.0453 & 0.0495 & 0.0513  & 0.0671             & 0.0642             & \underline{0.0685} & \textbf{0.0745}  &8.76\%\\
			                           & MovieLens100K & 0.0325 & 0.0353 & 0.0379  & 0.0448             & 0.0426             & \underline{0.0468} & \textbf{0.0498} &6.62\%\\
			                           & Digital 	   & 0.0422 & 0.0455 & 0.0570  & 0.0574             & 0.0589             & \underline{0.0641} & \textbf{0.0691}&7.64\% \\
			                           & Software	   & 0.0401 & 0.0428 & 0.0524  & 0.0532             & 0.0510             & \underline{0.0545} & \textbf{0.0580}&6.42\% \\
			\bottomrule
		\end{tabular}
	\end{center}
\end{table*}

\subsection{RQ1:Hyperparameter and Ablation Analysis}
In order to answer RQ1, we design a series of experiments
aiming to figure out the effect of the crucial hyperparameters
which will affect the performance of our model. We do the
ablation analysis as well which aims to investigate the effectiveness of the item trend information. Item trend information
plays a crucial role in our approach and distinguishes our
method from other sequential recommendation methods. There
are two factors affecting the performance of our proposed
methods: the first is the modeling of item trend information,
and the other factor is how the item trend information is
absorbed with the item embedding.

The hyperparameter \(q\) directly influence the first factor, and the hyperparameter \(\omega, \beta\), and\(\alpha\) significantly affect upon the second factor.
we study the impact of these two factors as follows.
\newline
\textbf{Item Trend Information Modeling}\hspace{0.2cm}
\begin{figure}[htbp]
	\centerline{\includegraphics[width=\linewidth]{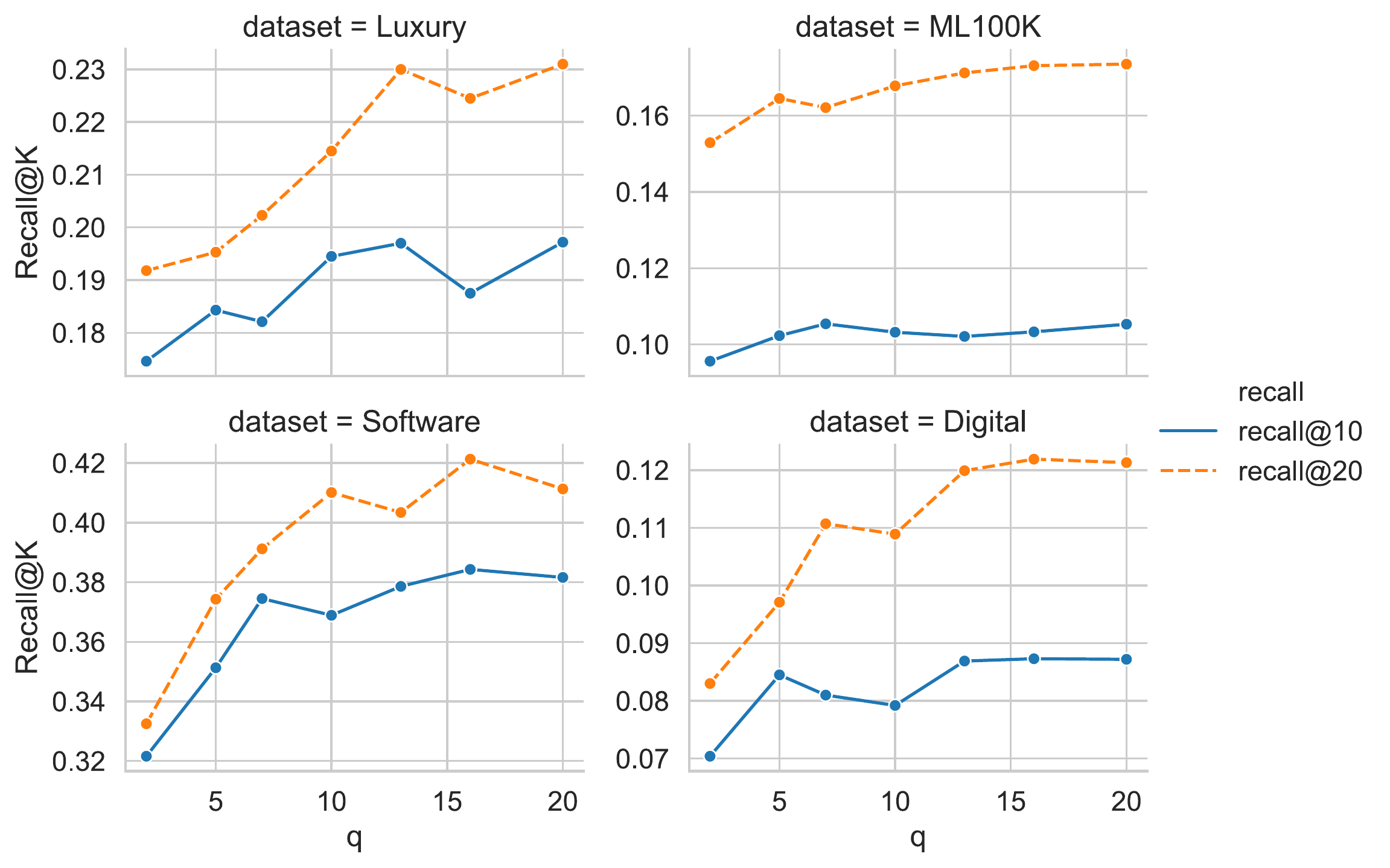}}
	\caption{Effect of \(q\) for item trend information}
	\label{q}
\end{figure}
The sequence length \(q\) of the recent user defines the time span of the trend for
an item. Based on the result shown in Fig.\ref{q}, we have following observations:
\begin{itemize}
	\item The performance goes slightly better when the length parameter $q$ is set to a higher value.
	\item When the value of \(q\) reaches a certain level,the result of performance goes stable.
	\item The optimal value of $q$ varies and depends on the dataset we choose. The datasets in which items have longer user interaction history require a relatively larger value of $q$, therefore the item trend is well represented. For example, a mechanical keyboard has gone out of fashion since 2000, but remains popular in the niche market. As such, one or two recent purchase record can not represent its falling trend.
\end{itemize}
\textbf{Sensitivity Analysis: \(\omega,\alpha \) and \(\beta \)}\hspace{0.2cm}
The proportion parameters determine how much percentage the item trend information should be considered, and thus affect the final prediction score. To be more concretely, \(\omega\) indicates the proportion of the long-term
user preference, and correspondingly \(1-\omega\) reflects the proportion of the short-term user preference.
Similarly, \(\alpha\) and \(\beta\) are the proportions of the item trend information.
\begin{figure}[htbp]
	\centerline{\includegraphics[width=\linewidth]{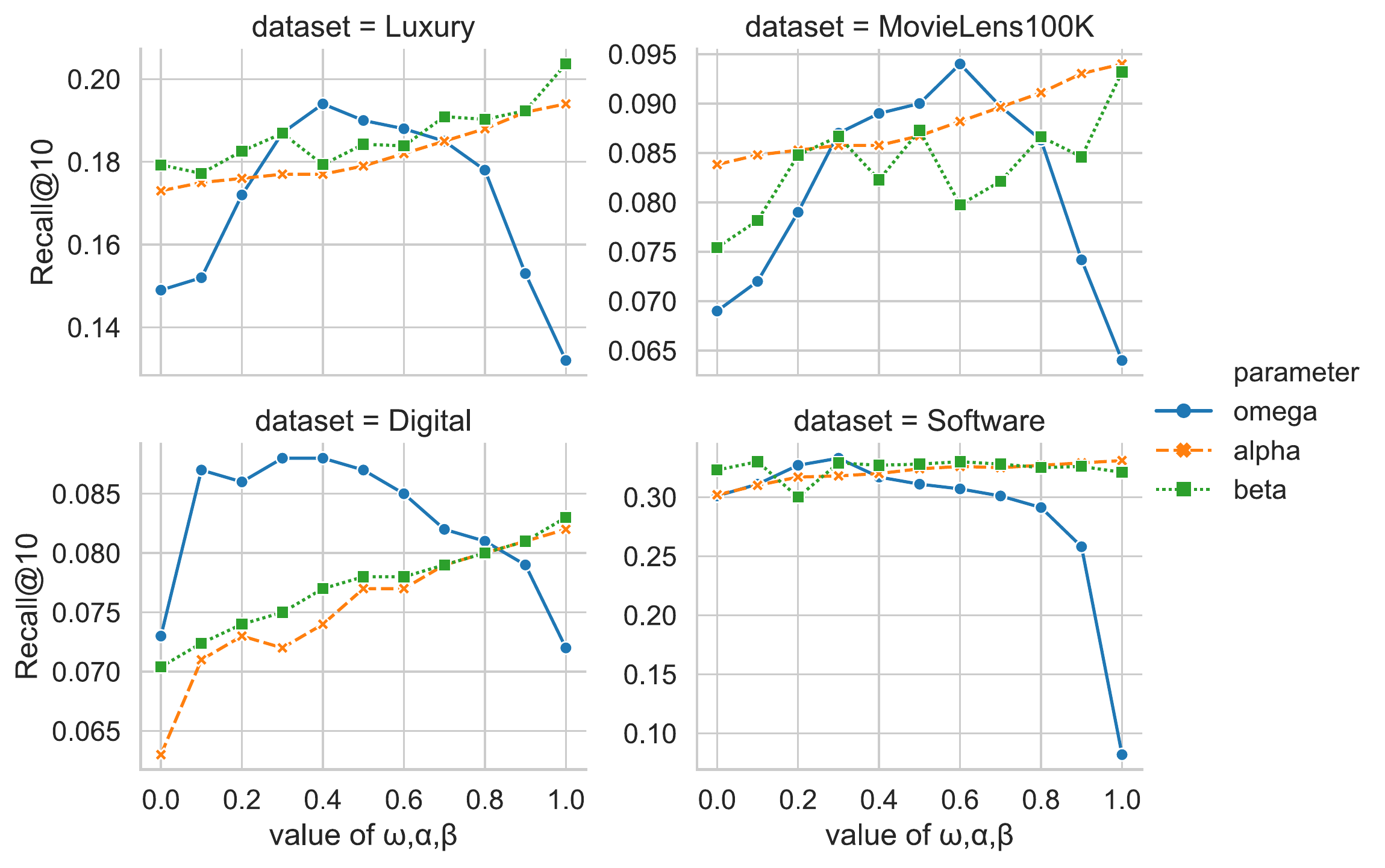}}
	\caption{Effects of the hyper-parameters: \(\omega,\alpha,\beta\) on luxury and MovieLens datasets}
	\label{omega}
\end{figure}

\begin{figure}[htbp]
	\centerline{\includegraphics[width=\linewidth]{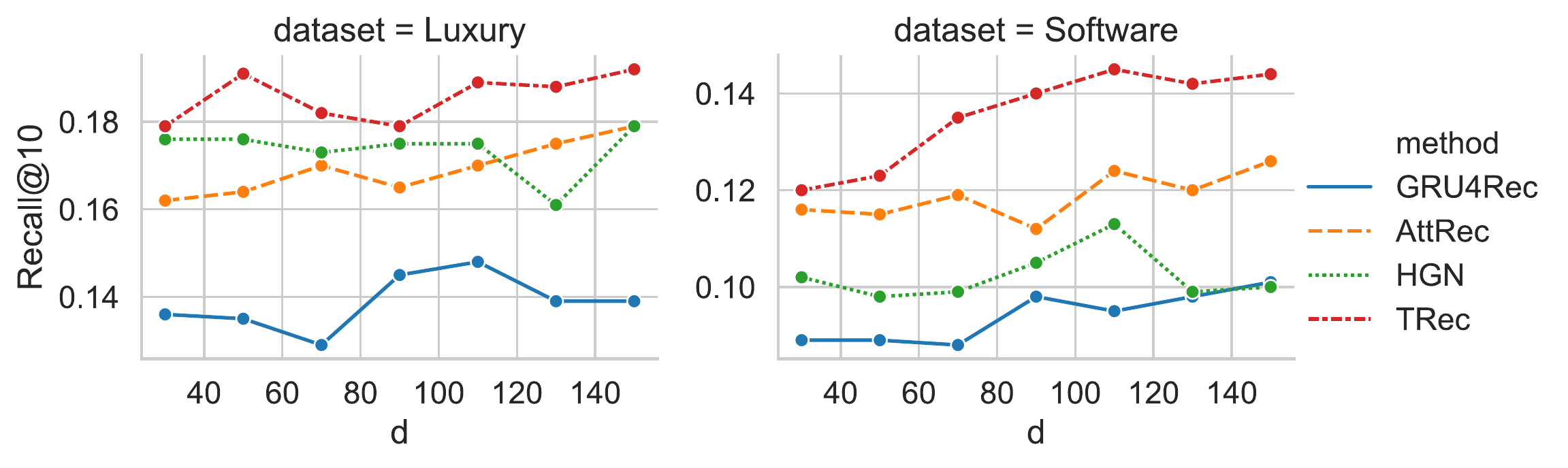}}
	\caption{Effects of dimension of embedding: \(d\)}
	\label{dimension}
\end{figure}

The impact of proportion parameters is shown in Fig. \ref{omega}. 
The optimal value of these parameters varies when different datasets are adopted. 
As we can obviously see from this figure that the higher values of \(\alpha\) and \(\beta\) result in the better performance on the four datasets. 
We can also find that the optimal value of \(\omega\) varies on four datasets, however, the performance downgrade rapidly as \(\omega\) goes toward 1. This reveals the importance of user short-term preference.
\newline
\textbf{Dimension of Embedding}\hspace{0.2cm} Embedding vector carries the information of users and items. The dimension of embedding vector dictates the capacity of information of a user or an item. We test the effect of varying dimension lengths on two datasets: Amazon-Luxury and Software.
We present the effect of dimension of embedding \(d\) on Recall@10 in Fig. \ref{dimension}. It shows that embedding with a dimension less than 40 is incapable to represent enough information of a user or an item. As \(d\) grows
larger, the performance goes up with an unstable fluctuation. Generally speaking, the performance of all the models behaves much better when \(d\) is larger than 100.
\\
\textbf{Ablation Analysis}\hspace{0.2cm} In order to answer RQ1 and find out the effectiveness of different components in our proposed framework, to this end, we take apart our model and evaluate the performance with part of its component taken away.
The results are based on Amazon-Luxury and MovieLens100K. Recall@10 and NDCG@10 are used as the evaluating metrics for our test. R@10 in Table IV stands for Recall@10.
The results shown in Table \ref{ablation} demonstrate the effect of these different model variations.
The default setting achieves the ideal marks over the Luxury and ML100K two datasets. We can see our model with the self-attention layer taken away decreased in performance. The reason might be that the self-attention layer efficiently captures the interrelation of the variables within the features of user and item. We also notice the impact of removing item trend information comes with a decrease in performance. The Amazon-Luxury dataset have a larger rate of performance decreasing, which is supposed to attribute to a shorter time-span of the item trend.
(5) shows the user short-term preference plays an important role in both datasets. Removing the user short-term representation leads to a consistent performance degradation in our test. (6) shows a significant loss in performance. Compared to (5) which suffers a relatively modest decrease, it seems that the item trend information makes a compensation on the accuracy in some way. 

We also experiment with different attentive matrix aggregation methods. In the default situation, the average
method is adopted. We test the max aggregation as an alternative way. The result in (7-8) shows a slight fluctuation in performance. So we conclude that the impact of aggregation is related to different datasets.
\begin{table}[htbp]
	\begin{center}
		\caption{Ablation analysis, w/ and w/o stand for with and without,iti stand for Item trend information}
		\label{ablation}
		\begin{tabular}{l|cccc}
			\toprule
			\multirow{2}{*}{Architecture} & \multicolumn{2}{c}{Amazon-Luxury} & \multicolumn{2}{c}{MovieLens100K}                                   \\
			                              & R@10                              & NDCG@10                           & R@10           & NDCG@10        \\
			\midrule
			(1) TRec                      & \textbf{0.194}                    & \textbf{0.096}                    & \textbf{0.103} & \textbf{0.096} \\
			(2) w/o Self-Att              & 0.182                             & 0.085                             & 0.095          & 0.083          \\
			(3) w/o ITI                   & 0.174                             & 0.078                             & 0.093          & 0.079          \\
			(4) w/o Self-Att\&ITI         & 0.170                             & 0.074                             & 0.089          & 0.075          \\
			(5) w/o u+                    & 0.179                             & 0.081                             & 0.094          & 0.081          \\
			(6) w/o u+\&ITI               & 0.132                             & 0.051                             & 0.064          & 0.051          \\
			(7) aggr-avg                  & 0.194                             & 0.096                             & 0.103          & 0.096          \\
			(8) aggr-max                  & 0.192                             & 0.094                             & 0.103          & 0.095          \\
			\bottomrule
		\end{tabular}
	\end{center}
\end{table}

\subsection{RQ2:Performance Comparison}
The performance comparison result is shown in Table \ref{performance}. The best performance is boldfaced while the scores underlined comes to a second place in performance. It is demonstrated that our proposed model TRec achieves the best performance
among all the datasets. The superior performance answers RQ2. By integrating the item trend information and the user short-term preference together with the long-term user and item representation, our model has significantly enhanced the effectiveness of sequential recommendation tasks. This table also obviously shows the listed non-neural approaches (i.e. MF and FPMC), are significantly outperformed by our method with an up to 10 percent gain. A intuitive explanation is that these latent factor based models view users and items as being static. Hence, they neglects the potential and latent information within them, such as the item trend information and the user short-term preference. Comparing to the latest neural-based methods, our model achieves an encouraging gain in performance. We can see that our proposed model exhibits the improvements over other baseline models from 6\% to 10\%
on different datasets. It is noteworthy that our model achieves a more competitive gain on the Luxury dataset. A reasonable explanation is that the luxury item goes out of fashion in a relatively shorter time span.

\addtolength{\intextsep}{-3pt}
\begin{table}[htbp]
	\begin{center}
		\caption{Training time cost comparison(per epoch) in terms of seconds}
		\label{timecost}
		\begin{tabular}{c|ccccc}
			\toprule
			Dataset       & GRU4Rec & Caser & AttRec & HGN & TRec \\
			\midrule
			MovieLens100K & 16      & 14    & \textbf{1.8}    & 2.5 & 1.9  \\
			Luxury		  & 40      & 32    & 7.8    & 9.0 & \textbf{7.8}  \\
			Digital 		  & 30      & 23    & \textbf{5.7}    & 6.5 & 5.8  \\
			Software 	  & 33      & 28    & \textbf{6.5}    & 7.3 & 6.6  \\
			\bottomrule
		\end{tabular}
	\end{center}
\end{table}

\subsection{Training Efficiency}
We conduct the training efficiency test over four different benchmark datasets. Factorization based models are not tested
in this comparison as their recommendation accuracy is outperformed by the state-of-the-art models. All the tests are running over 100 epochs and the average time cost per epoch is recorded for evaluation. Table \ref{timecost} shows the deep learning based models like GRU4Rec \cite{Hidasi2016} and Caser \cite{Tang2018} cost a huge amount of time for training each epoch.
Our model achieves a significantly lower runtime cost over other neural-based baseline models. Therefore, our proposed method has a promising potential to be employed into the real industrial practice.

\section{Conclusion and Future Work}
In this paper, we propose a novel sequential recommendation approach TRec, 
which incorporates the item trend information into the consideration for the next item prediction. 
To the best of our knowledge, 
TRec is the first method that exploits the item trend information for prediction in the context of sequential recommendation. 
Extensive empirical results show that our proposed model achieves the superior performance versus the state-of-the-art models, 
while the runtime cost per epoch is smaller than the CNN-based or RNN-based approaches by a wide margin.

In the future, we would investigate a more efficient and precisely way in terms of representing the item trend information other than using the latest top k interacted users' embedding.
as well as combining the rich side information, such as user review. 
We would also investigate taking advantage of reinforcement learning to address the varying item trend issue.
Our model suffers another drawback that graph-based scenario like social network is not suitable for matrix-like modelling. 
Seeking method for item trend information representation in graph scenario might be another research direction.

\vspace{12pt}
\bibliography{MyCollection}

\begin{thebibliography}{10}
\providecommand{\url}[1]{#1}
\csname url@samestyle\endcsname
\providecommand{\newblock}{\relax}
\providecommand{\bibinfo}[2]{#2}
\providecommand{\BIBentrySTDinterwordspacing}{\spaceskip=0pt\relax}
\providecommand{\BIBentryALTinterwordstretchfactor}{4}
\providecommand{\BIBentryALTinterwordspacing}{\spaceskip=\fontdimen2\font plus
\BIBentryALTinterwordstretchfactor\fontdimen3\font minus
  \fontdimen4\font\relax}
\providecommand{\BIBforeignlanguage}[2]{{%
\expandafter\ifx\csname l@#1\endcsname\relax
\typeout{** WARNING: IEEEtran.bst: No hyphenation pattern has been}%
\typeout{** loaded for the language `#1'. Using the pattern for}%
\typeout{** the default language instead.}%
\else
\language=\csname l@#1\endcsname
\fi
#2}}
\providecommand{\BIBdecl}{\relax}
\BIBdecl

\bibitem{FANG2019}
H.~Fang, Z.~Danning, and S.~Yiheng, ``{Deep Learning for Sequential
  Recommendation: Algorithms, Influential Factors, and Evaluations},'' in
  \emph{SIGIR 2019 - Proceedings of the 42nd International ACM SIGIR Conference
  on Research and Development in Information Retrieval}.\hskip 1em plus 0.5em
  minus 0.4em\relax Association for Computing Machinery, Inc, jul 2019, pp.
  1281--1284.

\bibitem{Sarwar2001}
B.~Sarwar, G.~Karypis, J.~Konstan, and J.~Riedl, ``{Item-based collaborative
  filtering recommendation algorithms},'' in \emph{Proceedings of the 10th
  International Conference on World Wide Web, WWW 2001}.\hskip 1em plus 0.5em
  minus 0.4em\relax Association for Computing Machinery, Inc, apr 2001, pp.
  285--295.

\bibitem{Pazzani2007}
M.~J. Pazzani and D.~Billsus, ``{Content-based recommendation systems},'' in
  \emph{Lecture Notes in Computer Science (including subseries Lecture Notes in
  Artificial Intelligence and Lecture Notes in Bioinformatics)}, vol. 4321
  LNCS, 2007, pp. 325--341.

\bibitem{Koren2015}
M.~Hayakawa, ``{MF Techniques},'' in \emph{Earthquake Prediction with Radio
  Techniques}, 2015, pp. 199--207.

\bibitem{Rendle2010}
\BIBentryALTinterwordspacing
S.~Rendle, ``{Factorization machines},'' in \emph{Proceedings - IEEE
  International Conference on Data Mining, ICDM}, 2010, pp. 995--1000.
  [Online]. Available: \url{http://www.libfm.org}
\BIBentrySTDinterwordspacing

\bibitem{Jannach2015}
D.~Jannach, L.~Lerche, and M.~Jugovac, ``{Adaptation and evaluation of
  recommendations for short-term shopping goals},'' in \emph{RecSys 2015 -
  Proceedings of the 9th ACM Conference on Recommender Systems}.\hskip 1em plus
  0.5em minus 0.4em\relax Association for Computing Machinery, Inc, sep 2015,
  pp. 211--218.

\bibitem{Zhang2019}
S.~Zhang, L.~Yao, A.~Sun, and Y.~Tay, ``{Deep learning based recommender
  system: A survey and new perspectives},'' \emph{ACM Computing Surveys},
  vol.~52, no.~1, 2019.

\bibitem{Zhang2018}
\BIBentryALTinterwordspacing
S.~Zhang, Y.~Tay, L.~Yao, and A.~Sun, ``{Next Item Recommendation with
  Self-Attention},'' in \emph{arXiv}, aug 2018. [Online]. Available:
  \url{http://arxiv.org/abs/1808.06414}
\BIBentrySTDinterwordspacing

\bibitem{Fu2001}
P.~Fu, ``{The Cinema of Hong Kong: History, Arts, Identity (review)},''
  \emph{Asian Theatre Journal}, vol.~18, no.~2, pp. 279--281, 2001.

\bibitem{Davidson2010}
\BIBentryALTinterwordspacing
J.~Davidson, B.~Liebald, J.~Liu, P.~Nandy, and T.~{Van Vleet}, ``{The YouTube
  video recommendation system},'' in \emph{RecSys'10 - Proceedings of the 4th
  ACM Conference on Recommender Systems}, 2010, pp. 293--296. [Online].
  Available: \url{www.youtube.com/videos.}
\BIBentrySTDinterwordspacing

\bibitem{Jannach}
\BIBentryALTinterwordspacing
D.~Jannach and M.~Ludewig, ``{When recurrent neural networks meet the
  neighborhood for session-based recommendation},'' \emph{RecSys 2017 -
  Proceedings of the 11th ACM Conference on Recommender Systems}, pp. 306--310,
  2017. [Online]. Available: \url{http://dx.doi.org/10.1145/3109859.3109872}
\BIBentrySTDinterwordspacing

\bibitem{He2017}
\BIBentryALTinterwordspacing
R.~He and J.~McAuley, ``{Fusing similarity models with markov chains for sparse
  sequential recommendation},'' in \emph{Proceedings - IEEE International
  Conference on Data Mining, ICDM}, 2017, pp. 191--200. [Online]. Available:
  \url{https://sites.google.com/a/}
\BIBentrySTDinterwordspacing

\bibitem{Hidasi2016}
\BIBentryALTinterwordspacing
B.~Hidasi, A.~Karatzoglou, L.~Baltrunas, and D.~Tikk, ``{Session-based
  recommendations with recurrent neural networks},'' in \emph{4th International
  Conference on Learning Representations, ICLR 2016 - Conference Track
  Proceedings}, nov 2016. [Online]. Available:
  \url{http://arxiv.org/abs/1511.06939}
\BIBentrySTDinterwordspacing

\bibitem{Tang2018}
J.~Tang and K.~Wang, ``{Personalized top-N sequential recommendation via
  convolutional sequence embedding},'' in \emph{WSDM 2018 - Proceedings of the
  11th ACM International Conference on Web Search and Data Mining}, vol.
  2018-Febua, 2018, pp. 565--573.

\bibitem{Yuan2019}
\BIBentryALTinterwordspacing
F.~Yuan, A.~Karatzoglou, I.~Arapakis, J.~M. Jose, and X.~He, ``{A simple
  convolutional generative network for next item recommendation},'' in
  \emph{WSDM 2019 - Proceedings of the 12th ACM International Conference on Web
  Search and Data Mining}, vol.~19, 2019, pp. 582--590. [Online]. Available:
  \url{https://doi.org/10.1145/3289600.3290975}
\BIBentrySTDinterwordspacing

\bibitem{Liu2018a}
\BIBentryALTinterwordspacing
Q.~Liu, R.~Mokhosi, Y.~Zeng, and H.~Zhang, ``{STAMP: Short-term
  attention/memory priority model for session-based recommendation},''
  \emph{Proceedings of the ACM SIGKDD International Conference on Knowledge
  Discovery and Data Mining}, pp. 1831--1839, 2018. [Online]. Available:
  \url{https://doi.org/10.1145/3219819.3219950}
\BIBentrySTDinterwordspacing

\bibitem{Vaswani2017}
A.~Vaswani, N.~Shazeer, N.~Parmar, J.~Uszkoreit, L.~Jones, A.~N. Gomez,
  {\L}.~Kaiser, and I.~Polosukhin, ``{Attention is all you need},'' in
  \emph{Advances in Neural Information Processing Systems}, vol. 2017-Decem,
  2017, pp. 5999--6009.

\bibitem{Nair2010}
V.~Nair and G.~E. Hinton, ``{Rectified linear units improve Restricted
  Boltzmann machines},'' in \emph{ICML 2010 - Proceedings, 27th International
  Conference on Machine Learning}, 2010, pp. 807--814.

\bibitem{Rendle2009}
S.~Rendle, C.~Freudenthaler, Z.~Gantner, and L.~Schmidt-Thieme, ``{BPR:
  Bayesian personalized ranking from implicit feedback},'' in \emph{Proceedings
  of the 25th Conference on Uncertainty in Artificial Intelligence, UAI 2009},
  2009, pp. 452--461.

\bibitem{Ni}
J.~Ni, J.~Li, and J.~McAuley, ``{Justifying Recommendations using
  Distantly-Labeled Reviews and Fine-Grained Aspects},'' in \emph{arXiv}, 2019,
  pp. 188--197.

\bibitem{Ma2019}
\BIBentryALTinterwordspacing
C.~Ma, P.~Kang, and X.~Liu, ``{Hierarchical gating networks for sequential
  recommendation},'' in \emph{Proceedings of the ACM SIGKDD International
  Conference on Knowledge Discovery and Data Mining}, 2019, pp. 825--833.
  [Online]. Available: \url{https://doi.org/10.1145/3292500.3330984}
\BIBentrySTDinterwordspacing

\end{thebibliography}
\bibliographystyle{IEEEtran}
\end{document}